\documentclass[pra,twocolumn,showpacs]{revtex4}

\usepackage{graphicx}

\begin{document}
\title{Dissociation of Feshbach Molecules into Different Partial Waves}
\author{Stephan D{\"u}rr, Thomas Volz, Niels Syassen, and Gerhard Rempe}
\affiliation{Max-Planck-Institut f{\"u}r Quantenoptik, Hans-Kopfermann-Stra{\ss}e 1, 85748 Garching, Germany}
\author{Eric van Kempen, Servaas Kokkelmans, and Boudewijn Verhaar}
\affiliation{Eindhoven University of Technology, P.O.\ Box 513, 5600MB Eindhoven, The Netherlands}
\author{Harald Friedrich}
\affiliation{Physik-Department, Technische Universit\"at M\"unchen, 85748 Garching, Germany}
%
\hyphenation{Fesh-bach}
\begin{abstract}
Ultracold molecules can be associated from ultracold atoms by ramping the magnetic field through a Feshbach resonance. A reverse ramp dissociates the molecules. Under suitable conditions, more than one outgoing partial wave can be populated. A theoretical model for this process is discussed here in detail. The model reveals the connection between the dissociation and the theory of multichannel scattering resonances. In particular,  the decay rate, the branching ratio, and the relative phase between the partial waves can be predicted  from theory or extracted from experiment. The results are applicable to our recent experiment in $^{87}$Rb, which has a $d$-wave shape resonance.
\end{abstract}
%
\maketitle

\section{Introduction}
The association of ultracold molecules from ultracold atomic gases using Feshbach resonances was a major recent breakthrough in the field of cold molecules \cite{regal:03,herbig:03,duerr:04,strecker:03,cubizolles:03,jochim:03a,xu:03,zwierlein:04,thompson:05}. Experimentally, the method requires a slow magnetic-field ramp that crosses a Feshbach resonance in the proper direction. The molecules can be dissociated back into unbound atom pairs by ramping the magnetic field back through the Feshbach resonance. Initially, the atomic gas is so cold that only $s$-wave collisions are relevant in the gas. The molecular bound state that causes the Feshbach resonance is usually also an $s$-wave state, and the outgoing wave in the dissociation is again an $s$ wave.

If one chooses a Feshbach resonance where the molecular bound state is not an $s$-wave state, then one might wonder if outgoing waves other than the $s$ wave can be produced. Of course, if the outgoing wave is as cold as the incoming one, the $s$ wave will dominate again. But if the magnetic-field ramp for the dissociation is fast, then kinetic energy can be added during the dissociation \cite{mukaiyama:04,duerr:04a}. It thus seems feasible to populate outgoing higher partial waves. This prompts many questions: If one creates a higher partial wave, will there still be an $s$-wave component? Do the different partial waves form a coherent superposition or an incoherent mixture? What determines the relative phase and the branching ratio? And how fast is the dissociation process? Are certain Feshbach resonances better suited than others for creating a large fraction of a specific outgoing partial wave? Answering these questions is nontrivial and no theory has been developed on the subject, yet.

The key to a theoretical description of the dissociation process lies in the observation that the dissociation is ``half a collision". In a full collision, two atoms come together and then separate again. In the association and dissociation using Feshbach resonances, the experimenter can ``freeze" the population in the middle, after the atoms came together. He can even choose, how much time is spent between association and dissociation. Still, the association and dissociation can be regarded as the first and second half of one collision. The concept of a half collision proved useful in other contexts before (see e.g.\ Refs.~\cite{du:91,machholm:94,wells:01,schneider:03}). Unfortunately, the models developed there are not directly applicable here. The question is then how the dissociation can be linked quantitatively to scattering theory. The objective of the present paper is to establish this link.

The motivation for this investigation comes from an experiment we performed recently, where dissociation into two partial waves, $s$ and $d$, is observed \cite{volz:cond-mat/0410083}. The experiment employs $^{87}$Rb, where the dissociation is particularly interesting because of the presence of a $d$-wave shape resonance. Our theoretical studies in the present paper are geared towards explaining the results of this experiment. A very brief summary of the model was already presented in Ref.~\cite{volz:cond-mat/0410083}.

The paper is outlined as follows: Section \ref{sec-basics} begins with a brief summary of some basics of scattering theory. Section \ref{sec-resonance} presents the theory of scattering resonances for a single partial wave. The $d$-wave shape resonance in $^{87}$Rb is introduced as a specific example. In Sec.~\ref{sec-Feshbach}, magnetically tunable Feshbach resonances in the collision of ultracold atoms are introduced. Section \ref{sec-multichannel} describes some basics of scattering resonances with more than one partial wave and then discusses the combination of the shape resonance and the Feshbach resonance. The stage is then set for Sec.~\ref{sec-dissociation}, where the link between scattering theory and the dissociation of molecules is presented. The decay rate, branching ratio, and the relative phase between the partial waves as observed in the experiment of Ref.~\cite{volz:cond-mat/0410083} are explained with this model.

\section{Basics of Scattering Theory}
\label{sec-basics}

\subsection{General}
The problem of scattering two particles off one another is easily separated into center-of-mass and relative coordinates. The center-of-mass motion is trivial so that the problem is in the relative motion. The latter is equivalent to the scattering of one particle with the reduced mass $m_{red}$ off a potential $V(\vec r)$. The relative motion is characterized by the wave vector $\vec k$. The corresponding kinetic energy is $E=\hbar^2 k^2/(2m_{red})$.

Scattering theory is usually formulated as a time-independent process with an incoming plane wave with $\vec k$ pointing along the $z$-direction. One can show that the scattered wave at large radius falls off radially like a spherical wave. Hence, the scattering wave function has the asymptotic form
\begin{eqnarray}
 \label{eq-plane-wave}
\psi^{(+)}(\vec r)  \stackrel{r \rightarrow \infty}{\sim}
 e^{ikz} + f(\vartheta,\varphi) \frac{e^{ikr}}{r} \; ,
\end{eqnarray}
where $r,\vartheta,\varphi$ are spherical coordinates. The scattering amplitude $f(\vartheta,\varphi)$ is related to the differential scattering cross section
\begin{eqnarray}
\frac{d\sigma}{d\Omega} = | f(\vartheta,\varphi) |^2 \; ,
\end{eqnarray}
where $d\Omega = \sin \vartheta \, d\vartheta \, d\varphi$ is the differential solid angle. Finally, the total cross section $\sigma$ is obtained by integration of the differential cross section over the full solid angle. The task in scattering theory is to determine $f(\vartheta,\varphi)$ for a given potential $V(\vec r)$.

\subsection{Partial Waves}
It is often useful to expand the scattering problem in terms of partial waves, i.e.\ spherical harmonics $Y_{l,m_l}(\vartheta,\varphi)$. The result for the incoming plane wave is
\begin{eqnarray}
e^{ikz} = \sum_{l=0}^\infty i^l j_l(kr) Y_{l0}(\vartheta) \sqrt{4\pi(2l+1)} \; ,
\end{eqnarray}
where $j_l$ denotes the spherical Bessel function of order $l$. The outgoing wave is also written as a sum of partial waves. In this paper, we restrict the scattering problem to the case, where all outgoing partial waves have $m_l=0$. This is the case, e.g., if the potential is invariant under rotations around the $z$-axis. The experiment in Ref.~\cite{volz:cond-mat/0410083} does not have such a cylindrically symmetric potential, yet all outgoing partial waves have $m_l=0$, as we will see in Sec.~\ref{sec-two-resonances}. Hence,
\begin{eqnarray}
f(\vartheta) = \sum_{l=0}^\infty f_l Y_{l0}(\vartheta) \sqrt{4\pi(2l+1)} \; ,
\end{eqnarray}
where the partial-wave coefficients are labeled $f_l$. With these expansions, one can reformulate the scattering problem. As a first step, one solves the scattering problem for one incoming partial wave $l'$, where the asymptotic form of the scattering state is
\begin{eqnarray}
 \label{eq-define-S-matrix}
\psi^{(+)}_{l'}(\vec r) & \stackrel{r \rightarrow \infty}{\sim} &
(-1)^{l'} \frac{e^{-ikr}}{r} \; Y_{l'0} (\vartheta) \nonumber \\
&& - \frac{e^{ikr}}{r} \; \sum_{l=0}^\infty S_{ll'} Y_{l0} (\vartheta) \; .
\end{eqnarray}
The outgoing partial waves have certain complex amplitudes. These amplitudes form the so-called $S$-matrix (or scattering matrix). Conservation of the number of particles implies that the $S$-matrix is unitary. In addition, realistic Hamiltonians in atomic physics are invariant under time reversal, which implies that the $S$-matrix is symmetric.

The second step is to superpose the incoming partial waves with suitable amplitudes to obtain an incoming plane wave. This yields
\begin{eqnarray}
\label{eq-f-from-S}
f_l  = \frac{1}{2ik} \sum_{l'=0}^\infty \sqrt{\frac{2l'+1}{2l+1}} \left( S_{ll'} - \delta_{ll'}\right) \; ,
\end{eqnarray}
where $\delta_{ll'}$ is the Kronecker symbol. The task of calculating $f(\vartheta)$ for a given potential $V(\vec r)$ is therefore replaced by the task of calculating the $S$-matrix. Realistic potentials lead to selection rules for the angular momentum. Therefore, there are usually only few non-vanishing matrix elements in $S$. Calculating the $S$-matrix is therefore often easier than calculating $f(\vartheta)$ directly.

If more than one scattered partial wave is populated, then the differential cross section shows a spatial interference pattern between the partial waves. The {\em total} cross section, however, shows no interference because $\int Y^*_{l'0} Y^{ }_{l0} d\Omega =\delta_{ll'}$. Hence,
\begin{eqnarray}
\label{eq-sigma-from-fl}
\sigma = \sum_{l=0}^\infty \sigma_l = \sum_{l=0}^\infty 4\pi (2l+1) |f_l|^2 \; .
\end{eqnarray}
When calculating differential or total cross sections, special attention must be paid in the case of indistinguishable particles, because then the two-particle wave function needs proper symmetrization. We restrict the rest of this paper to the scattering of identical bosons, where the cross sections double for the even partial waves and vanish for the odd partial waves.

\subsection{Spherical Symmetry}
Things simplify if the potential is spherically symmetric. The quantum number $l$ is then conserved, so that the $S$-matrix is diagonal. Combined with unitarity this implies $|S_{ll'}|=\delta_{ll'}$. All the information about the $S$-matrix is therefore in the phases of the diagonal elements and one defines the scattering phase $\delta_l$ for each partial wave by
\begin{eqnarray}
\label{eq-def-delta}
S_{ll}= e^{2i\delta_l} \; .
\end{eqnarray}
Note that the scattering phase is real and only defined modulo $\pi$. The connection between $f_l$ and the $S$-matrix simplifies to
\begin{eqnarray}
 \label{eq-fl-symmetric}
f_l = \frac{S_{ll}-1}{2ik} \; .
\end{eqnarray}
For identical bosons, the total cross section for the $l$-th partial wave is 
\begin{eqnarray}
\sigma_l = (2l+1) \frac{8\pi}{k^2} \sin^2 \delta_l
\end{eqnarray}
if $l$ is even; and 0 otherwise. Since $\delta_l$ is real, $\sigma_l$ has an upper bound
\begin{eqnarray}
\label{eq-unitarity-limit}
\sigma_l^{max} = (2l+1) \frac{8\pi}{k^2} \; ,
\end{eqnarray}
which is called the unitarity limit.

\subsection{Threshold Behavior}
 \label{sec-threshold}
The low-energy limit of the scattering properties is often important. For reasons discussed in Sec.~\ref{sec-Feshbach-general}, the energy where $k=0$ is called the dissociation threshold. The behavior near threshold can often be expressed in terms of simple power laws. This was first systematically investigated by Wigner \cite{wigner:48}. If the potential is spherically symmetric and has a long-range tail following a power law $V(r)\propto r^{-s}$, then one can show that (see p.~230 in Ref.~\cite{taylor:72})
\begin{eqnarray}
 \label{eq-threshold-law}
f_l \stackrel{k \rightarrow 0}{\sim} 
\left\{ 
\begin{array}{ll}
O(k^{2l}) & \mbox{if } \; 2l \leq s-3\\
O(k^{s-3}) & \mbox{otherwise} \; .
\end{array}
\right. 
\end{eqnarray}
Hence, Eqs.~(\ref{eq-sigma-from-fl})-(\ref{eq-fl-symmetric}) imply that $\delta_l$ vanishes near threshold like $k f_l$ and that $\sigma_l$ vanishes like $f_l^2$. For $s>3$ it follows that, in the low-energy limit, $s$-wave scattering dominates over all other partial waves and that $\delta_0 \stackrel{k \rightarrow 0}{\sim} O(k)$. This motivates the definition of the $s$-wave scattering length \cite{note:tan-delta}
\begin{eqnarray}
 \label{eq-scattering-length}
a = - \lim_{k\rightarrow 0} \frac{\delta_0}{k} \; .
\end{eqnarray}
The total scattering cross section for identical bosons is then
\begin{eqnarray}
\sigma \stackrel{k \rightarrow 0}{\sim} 8\pi a^2 \; .
\end{eqnarray}
The regime of cold collisions is characterized by energies that are so low that only few partial waves have a noticeable scattering cross section. If only $s$-waves are important, the collisions are called ultracold.

\subsection{Coupled-Channels Calculations}
After performing the partial-wave expansion, the remaining problem in scattering theory lies in the calculation of the $S$-matrix for a given potential. In essence, the Schr\"odinger equation must be solved along the radial coordinate. Various numerical methods have been developed to solve this problem. If more than one collision channel is involved, then coupling between the channels must be taken into account. The corresponding calculations are called coupled-channels calculations.

For atom-atom collisions (except for atomic hydrogen) there is another problem: ab-initio calculations for the interaction potentials are not accurate enough to make realistic predictions for the cold-collision properties. As a solution, some quantities, such as the van-der-Waals coefficient $C_6$, are treated as free fit parameters and experimental input is used to constrain the model, in order to obtain realistic predictions for the cold-collision properties (see e.g.\ Ref.~\cite{kempen:02}).

\section{Resonance Scattering}
\label{sec-resonance}

\subsection{S-Matrix}
Resonance scattering relies on the presence of a quasi-bound state. A quasi-bound state is a discrete state just like a bound state, but with an energy above threshold. Hence, if one prepares population in this state, it will undergo spontaneous exponential decay into unbound states. Examples for quasi-bound states are given in Secs.~\ref{sec-shape-resonance} and \ref{sec-Feshbach}. The decay of the quasi-bound state is a dissociation process, because a quasi-bound system decays into two unbound particles. The words decay and dissociation are therefore synonymous in the present context.

In a scattering experiment, some fraction of the incoming flux can make the transition to the quasi-bound state and subsequently decay back into unbound states. Obviously, the probability to make this transition must depend on the energy difference between the incoming flux and the quasi-bound state. When the energies match, the population of the quasi-bound state is resonantly enhanced, while far-off resonance the population of the quasi-bound state becomes negligible.

Near resonance, the $S$-matrix is changed drastically. In this section, we consider only the case, where the $S$-matrix is diagonal and only one partial wave has a resonance. With some effort one can show that the relevant $S$-matrix element is well approximated by a Breit-Wigner form \cite{breit:36,taylor:72} (see appendix \ref{sec-app-poles-of-S} for a derivation)
\begin{eqnarray}
 \label{eq-S-matrix-single-channel}
S_{ll} = e^{2i\delta_l^{bg}}\left( 1- \frac{i\hbar\Gamma}{E-E_{res}+i\hbar\Gamma/2} \right) \; .
\end{eqnarray}
Here $\delta_l^{bg}$ is the background value of the scattering phase for the relevant partial wave. This value is reached for energies far away from the resonance. $E_{res}$ is the energy, at which the resonance occurs. The parameter $\Gamma$ must be positive and can be interpreted as the decay rate of the quasi-bound state as discussed in Sec.~\ref{sec-pulsed-scattering}. Generally, $\delta_l^{bg}$ and $\Gamma$ can depend on $E$, but in the following we assume that they are independent of $E$ within the width of the resonance.

\begin{figure} [t!]
\includegraphics[width=.4\textwidth]{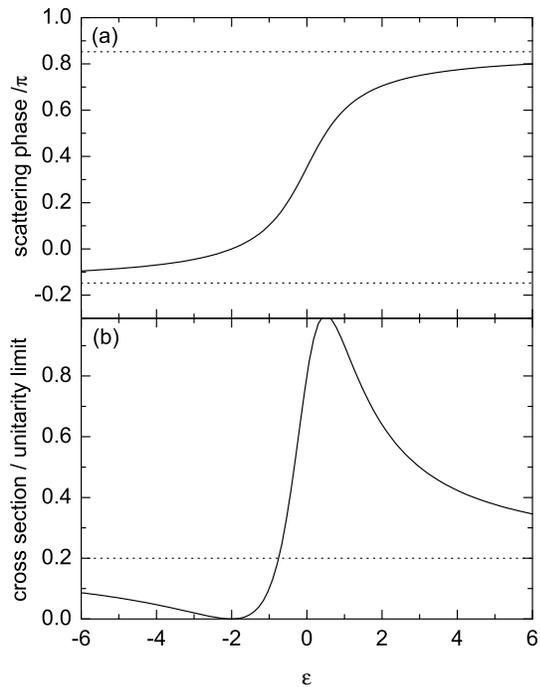}
\caption{\label{fig-Fano-profile}
A typical scattering resonance. (a) Scattering phase calculated from Eqs.~(\ref{eq-delta-l}) and (\ref{eq-delta-res}) for $q=-\cot \delta_l^{bg} =2$ as a function of the dimensionless energy $\epsilon$. The horizontal lines show $\delta_l^{bg}$ and $\delta_l^{bg}+\pi$, which are reached for $\epsilon \rightarrow \pm \infty$. (b) Beutler-Fano profile for the cross section calculated from Eq.~(\ref{eq-Fano}) for the same resonance. The horizontal line shows the background value.
 }
\end{figure}

It is customary to introduce the dimensionless detuning of the energy from resonance
\begin{eqnarray}
 \label{eq-def-epsilon}
\epsilon = \frac{2}{\hbar\Gamma} (E-E_{res}) \; .
\end{eqnarray}
Since $|S_{ll}|=1$, one can again write $S_{ll}=e^{2i\delta_l}$, where $\delta_l$ is real and one can easily show that
\begin{eqnarray}
 \label{eq-delta-l}
\delta_l = \delta_l^{bg} + \delta_l^{res}
\end{eqnarray}
with
\begin{eqnarray}
 \label{eq-delta-res}
\epsilon = - \cot \delta_l^{res} \; .
\end{eqnarray}
This means that $\delta_l(E)$ has the form of an inverse tangent plus an offset, as shown in Fig.~\ref{fig-Fano-profile}a. In particular, $\delta_l$ increases by $\pi$, as the energy moves all the way through resonance. Right on resonance $\delta_l^{res} = \pi/2$ and $\delta_l$ has an inflection point.

\subsection{Cross Section}
From the $S$-matrix in Eq.~(\ref{eq-S-matrix-single-channel}), one can easily derive an expression for the total cross section of the relevant partial wave, yielding
\begin{eqnarray}
 \label{eq-Fano}
\sigma_l = \sigma_l^{bg} \; \frac{(q+\epsilon)^2}{1+\epsilon^2}
\end{eqnarray}
with
\begin{eqnarray}
q = - \cot \delta_l^{bg}
\end{eqnarray}
and 
\begin{eqnarray}
\label{eq-def-sigma-bg}
\sigma_l^{bg}=(2l+1)\frac{8\pi}{k^2}\sin^2\delta_l^{bg} \; .
\end{eqnarray}
Equation (\ref{eq-Fano}) is called a Beutler-Fano profile. This asymmetric resonance profile was first observed experimentally by Beutler in an autoionization experiment \cite{beutler:35} and then explained by Fano \cite{fano:35,fano:61}. An example of a Beutler-Fano profile for $q=2$ is shown in Fig.~\ref{fig-Fano-profile}b.

A physical interpretation of the asymmetry in the Beutler-Fano profile is obtained easily when writing the $S$-matrix from Eq.~(\ref{eq-S-matrix-single-channel}) as the sum of a background scattered part $S_{ll}^{bg}=e^{2i\delta_l^{bg}}$ and a resonantly scattered part
\begin{eqnarray}
\label{eq-sum-S-matrix}
S_{ll} = S_{ll}^{bg} + S_{ll}^{res} \; .
\end{eqnarray}
Note that $S_{ll}^{res}$ is usually {\em not} unitary. It is merely the resonant contribution to the $S$-matrix, but not really an $S$-matrix by itself.

It is obvious from Eq.~(\ref{eq-S-matrix-single-channel}), that the phase of $S_{ll}^{res}$ changes by $\pi$, when $E$ moves all the way through resonance. Hence the interference between $S_{ll}^{bg}$ and $S_{ll}^{res}$ changes from constructive to destructive, or vice versa. It follows from Eq.~(\ref{eq-Fano}), that there is complete destructive interference at $\epsilon = -q$, while the unitarity limit Eq.~(\ref{eq-unitarity-limit}) is reached at $\epsilon = 1/q$. At these points $\delta_l$ reaches $0$ and $\pi/2$, respectively. The resonance position, i.e.\ $\epsilon=0$, is at the inflection point of $\delta_l$, which is usually not identical to the maximum of $\sigma_l$. Note that $q$ can be positive or negative, so that the region of destructive interference can occur on either side of the resonance. For $|q| \gg 1$ (i.e.\ $\delta_l^{bg} \sim 0$), the Beutler-Fano profile is well approximated by a Lorentzian (except way out in the wings).

\subsection{Decay of the Quasi-Bound State}
\label{sec-pulsed-scattering}
Scattering theory is usually formulated as a time-independent problem. But in order to obtain a physical interpretation of the parameter $\Gamma$ in Eq.~(\ref{eq-S-matrix-single-channel}), we now consider resonance scattering of a {\em pulsed} incoming wave. The incoming pulse can be expressed as a superposition of incoming plane waves, each of which can be described by time-independent scattering theory. We first consider the case where the energy width $\Delta E$ of the incoming pulse is large, i.e.\ $\Delta E \gg \hbar \Gamma$. In this case, one can show (see p.~254 in Ref.~\cite{taylor:72}) that the shape of the scattered wave packet has a tail at long times $t$. In the tail, the probability of detecting a scattered particle falls off like $e^{-\Gamma t}$. This suggests the following interpretation: Some fraction of the population makes the transition to the quasi-bound state and decays from there with a rate $\Gamma$ into outgoing waves. $\Gamma$ is thus interpreted as the decay rate of the quasi-bound state.

Since $\Delta E \gg \hbar \Gamma$, the minimum duration of the incoming pulse $\Delta t\sim \hbar/\Delta E$ is much shorter than the mean lifetime of the quasi-bound state $1/\Gamma$ and the tail can easily be distinguished from the background scattered wave packet. In the opposite limit, where $\Delta E \ll \hbar \Gamma$, the $S$-matrix is nearly constant within $\Delta E$ and the scattering process leaves the temporal shape of the pulse almost unchanged.

For time-independent elastic scattering, conservation of energy implies that the outgoing wave must have the same energy as the incoming wave. This is not so simple in resonance scattering of a short pulse. Here, the resonance plays the role of an energy filter. Only incoming energies with $|E-E_{res}| < \hbar \Gamma$ have a large probability to make the transition to the quasi-bound state. Only these energies are found in the exponential tail of the decaying wave function.

\subsection{Resonances near Threshold}
While the consideration of pulsed scattering yields an intuitive interpretation of the parameter $\Gamma$ in terms of a decay rate, one can show that for time-independent scattering the parameter $\Gamma$ in Eq.~(\ref{eq-S-matrix-single-channel}) can depend on the collision energy. This is because, $\Gamma$ is not merely a property of the quasi-bound state. Instead, it describes the decay into outgoing waves, so that the energy of the outgoing waves is important for $\Gamma$. We will now explain this in more detail.

\begin{figure} [t!]
\includegraphics[width=.4\textwidth]{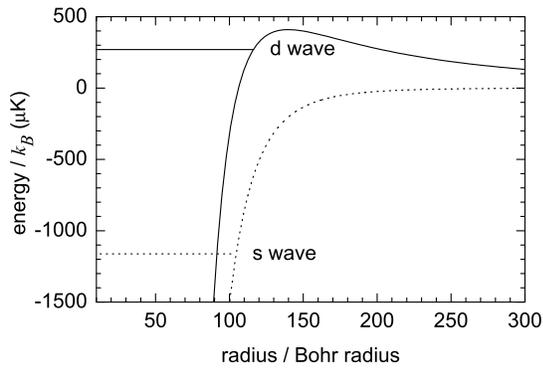}
\caption{\label{fig-potential-barrier}
A quasi-bound state in the collision of two $^{87}$Rb atoms. Potentials for the $s$ wave (dotted line) and the $d$ wave (solid line) are shown as a function of radius. The centrifugal barrier for the $d$ wave is clearly visible. In addition, the highest-lying $s$-wave bound state (horizontal dotted line) and the corresponding $d$-wave state (horizontal solid line) are shown. The $d$-wave state lies above threshold (i.e.\ $E>0$), so that it is only quasi-bound and gives rise to a scattering resonance.
 }
\end{figure}

The energy dependence of $\Gamma$ is particularly strong near threshold due to the centrifugal barrier. The centrifugal barrier arises from the centrifugal potential 
\begin{eqnarray}
V_{centr}= \frac{\hbar^2 l(l+1)}{2m_{red}\,r^2} \; ,
\end{eqnarray}
which depends on $l$ and vanishes for the $s$ wave. Competing with the van-der-Waals potential $-C_6 r^{-6}$, the centrifugal potential dominates at large radius and the van-der-Waals potential dominates at shorter radius. This results in a centrifugal barrier of finite height, as shown in Fig.~\ref{fig-potential-barrier}. Since the centrifugal potential vanishes at large radius, it does not shift the threshold energy.

With the centrifugal barrier, the energy dependence of $\Gamma$ near threshold can be understood as follows: For energies below the height of the centrifugal barrier, population in a quasi-bound state must tunnel through the centrifugal barrier to decay into an outgoing wave. When the incoming energy approaches threshold, so does the outgoing energy because energy is conserved in time-independent elastic scattering. Thus when approaching threshold, the distance through which the particles must tunnel diverges. We thus expect that $\Gamma$ vanishes near threshold, at least for $l\neq0$.

A more quantitative way to show that $\Gamma$ must vanish near threshold is the following: When inserting the $S$-matrix Eq.~(\ref{eq-S-matrix-single-channel}) into Eq.~(\ref{eq-fl-symmetric}), one finds that $f_l$ cannot vanish faster than $\Gamma/k$. Equation (\ref{eq-threshold-law}) for $f_l$ therefore sets a limit on how fast $\Gamma$ must vanish. One can show that the following threshold law applies \cite{mies:00a}
\begin{eqnarray}
 \label{eq-threshold-Gamma}
\Gamma \stackrel{k \rightarrow 0}{\sim} O(k^{2l+1}) \; .
\end{eqnarray}
This is valid for all $l$, even if the potential has a long-range tail following a power law $V(r)\propto r^{-s}$ \cite{mies:00a}. Hence, for $2l>s-3$ we find that $\Gamma$ vanishes even faster than $k f_l$. For the decay into $s$-waves, the physical meaning of this threshold law arises from the density of final states, which is proportional to $k$ (see e.g.\ Refs.~\cite{mukaiyama:04,timmermans:99}).

The Beutler-Fano profile for the cross section is typically distorted for resonances near threshold by the energy dependence of $\delta_l^{bg}$ and $\Gamma$, as well as by the factor $k^{-2}$ in the unitarity limit Eq.~(\ref{eq-unitarity-limit}). $\delta_l^{bg}$ follows the usual threshold law $\delta_l^{bg} \stackrel{k\rightarrow 0}{\sim} k f_l$ with $f_l$ following Eq.~(\ref{eq-threshold-law}). Note that for the molecule dissociation experiment in Sec.~\ref{sec-dissociation}, $k$ in Eq.~(\ref{eq-threshold-Gamma}) is the wave vector of the {\em outgoing} wave.

\subsection{Shape Resonance in $^{87}$Rb}
 \label{sec-shape-resonance}
A shape resonance is a scattering resonance, which is caused by a quasi-bound state behind some potential barrier. In this section we discuss a specific example, namely the $d$-wave shape resonance in the collision of two cold $^{87}$Rb atoms. We assume that both atoms are initially prepared in the lowest hyperfine state $|f,m_f\rangle = |1,1\rangle$ of the electronic ground state. This two-atom system has a quasi-bound $d$-wave state localized behind the centrifugal barrier as shown in Fig.~\ref{fig-potential-barrier}. Also shown in Fig.~\ref{fig-potential-barrier} are the $s$-wave and $d$-wave potentials and the highest-lying $s$-wave bound state. The centrifugal potential raises the energy of the corresponding $d$-wave state. In $^{87}$Rb the energy of this state lies below the top of the centrifugal barrier, so that it is not pushed out into the unconstrained continuum. But at the same time it lies above threshold and can therefore decay into the continuum by tunneling through the centrifugal barrier, so that it is a quasi-bound state. Tunneling also allows incoming flux in a scattering experiment to populate the state. Since the terms in the interaction Hamiltonian that can change $l$ are weak (see Sec.~\ref{sec-selection-rules}), this quasi-bound $d$-wave state couples almost exclusively to the $d$-wave continuum, so that a shape resonance is caused for incoming $d$ waves, but no noticeable resonance is caused for incoming $s$ waves.

\begin{figure} [t!]
\includegraphics[width=.4\textwidth]{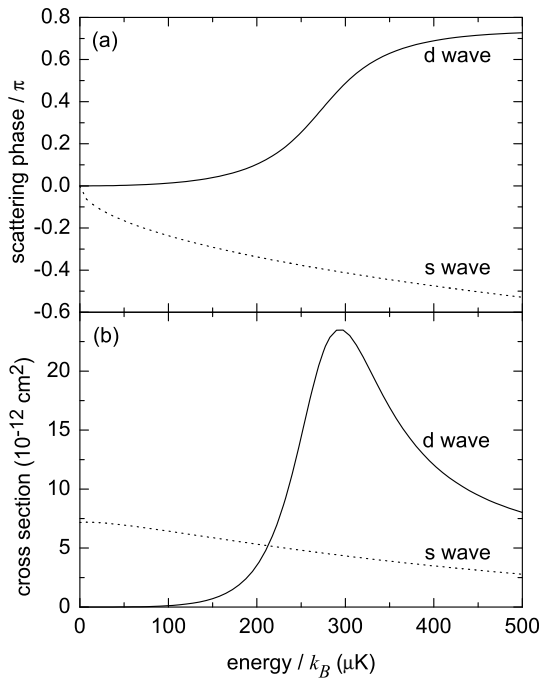}
\caption{\label{fig-87Rb-shape-resonance}
Shape resonance for the scattering of two $^{87}$Rb atoms in the hyperfine state $|1,1\rangle$ at $B=0$. (a) Scattering phases and (b) cross sections are shown for the $s$ wave (dotted lines) and the $d$-wave (solid lines). The $d$-wave shape resonance is clearly visible. Other partial waves yield no noticeable scattering in this energy range.
 }
\end{figure}

The partial-wave scattering phases and cross sections determined from a coupled-channels calculation are shown in Fig.~\ref{fig-87Rb-shape-resonance} for zero magnetic field, $B=0$. The threshold law Eq.~(\ref{eq-threshold-law}) for the $s$ wave predicts that $\delta_0 \stackrel{E \rightarrow 0}{\sim} O(E^{1/2})$ and that $\sigma_0$ approaches some finite value for small $E$. This is the case for small values of $E$, as is clearly seen in Fig.~\ref{fig-87Rb-shape-resonance}.

The $d$-wave scattering phase and the $d$-wave cross section both show clear signatures of the shape resonance. In contrast to the general discussion above, $\delta_2^{bg}$ decreases considerably over the width of the resonance. As a consequence, $\delta_2$ increases by noticeably less than $\pi$. From the threshold law Eq.~(\ref{eq-threshold-law}) with $s=6$, one expects $\sigma_2^{bg} \stackrel{E \rightarrow 0}{\sim} O(E^3)$, which leads to a strong asymmetry of the resonance in the cross section merely from the change in the background value. The additional inherent asymmetry of the Beutler-Fano profile is small for this resonance because $|q| \gg 1$.

The position and width of the shape resonance can be read-off from Fig.~\ref{fig-87Rb-shape-resonance}a. The inflection point of $\delta_2(E)$ lies at $E_{shape} = k_B \times 273~\mu$K. According to Eq.~(\ref{eq-delta-res}), the slope at the inflection point is $d\delta_2/dE = 2/(\hbar \Gamma)$, which yields $\Gamma=17$~MHz.

While a coupled-channels calculation is needed to obtain Fig.~\ref{fig-87Rb-shape-resonance}, one can already make a rough estimate for $\Gamma$ if only $E_{shape}$ and $C_6=4707$~a.u.\ \cite{marte:02} are known (1~a.u.\ $= 9.573\times 10^{-80}$~Jm$^6$). To this end, we approximate the potential as $V=V_{centr}-C_6 r^{-6}$, which is a good approximation at large radius. With this, we first calculate the tunneling probability in the WKB approximation for a spherical wave at energy $E_{shape}$ and obtain 25\%. Alternatively, the tunneling probability can be calculated analytically using a near-threshold approximation as discussed in Ref.~\cite{moritz:01}. Equation~(57) in Ref.~\cite{moritz:01} yields 23\%. Second, we consider a {\em classical} particle with mass $m_{red}$ released at rest at the outer classical turning point of the quasi-bound state. In the potential $V=V_{centr}-C_6 r^{-6}$, this particle will roll down to $r=0$ in 6.3~ns. The approximative treatment of the potential is reasonable, because most of the time is spent near the outer turning point, where $V= V_{centr}-C_6 r^{-6}$ is a good approximation. We assume that the particle is simply reflected at $r=0$ so that the round-trip time is 12.6~ns. We thus obtain $\Gamma \sim 0.25/(12.6~{\rm ns})=20$~MHz, which is quite close to the above result.

All other hyperfine levels of the electronic ground state of $^{87}$Rb have a similar shape resonance with almost the same energy and lifetime. The first experimental observation of this quasi-bound state had been made in a photoassociation experiment \cite{boesten:97}. For the hyperfine state $|2,2\rangle$, this shape resonance was recently investigated in two scattering experiments \cite{thomas:04,buggle:04}. Other atomic species also have shape resonances in cold collisions, as measured e.g.\ in Refs.~\cite{boesten:96,elbs:99,demarco:99,cote:99,burke:99a,williams:99,weinstein:02}.

\section{Magnetically tunable Feshbach Resonances}
\label{sec-Feshbach}
\subsection{General}
\label{sec-Feshbach-general}
\begin{figure} [t!]
\includegraphics[width=.4\textwidth]{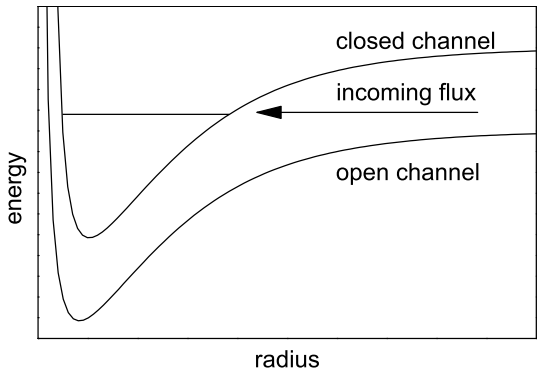}
\caption{\label{fig-Feshbach}
Scheme of a Feshbach resonance. The interaction Hamiltonian can cause transitions between the closed channel and the incoming flux in the open channel. A Feshbach resonance occurs when the energy of the incoming flux matches the energy of a closed-channel quasi-bound state.
 }
\end{figure}

In many scattering experiments, the particles have internal degrees of freedom, such as spin. For each spin state, there is a different potential. These potentials are referred to as the scattering channels. We consider the situation sketched in Fig.~\ref{fig-Feshbach}, where we picked two channels with different threshold energies. We assume that the incoming flux has the spin configuration of the lower channel and that the energy of the incoming flux is below the threshold of the upper channel. In this situation, no flux can emerge in the upper channel for energetic reasons. This channel is therefore energetically closed, while the entrance channel is always open. If the entrance channel is the only open channel, then a two-body collision cannot be inelastic, i.e.\ the spin states before and after scattering must be identical.

A Feshbach resonance arises if incoming flux in the open channel is resonant with the energy of a bound state in a closed channel. We call this bound state involved in the Feshbach resonance the ``molecular state". For the resonance to occur, the interaction Hamiltonian must be able to flip the spins, in order to cause transitions between the two channels. These spin flips lead to decay of the molecular state into unbound open-channel states, so that the molecular state is only quasi-bound.

The potentials shown in Fig.~\ref{fig-Feshbach} schematically represent Born-Oppenheimer potentials for the collision of two alkali atoms. The potentials have a long-range van-der-Waals tail, a deeply bound region due to the exchange interaction, and a repulsive part at very short radius due to the Coulomb interaction of the nuclei and a repulsive exchange interaction of the overlapping electron clouds. The energies in Fig.~\ref{fig-Feshbach} are not to scale. For $^{87}$Rb, for example, the dissociation energy of the singlet potential is $k_B \times 5750$~K \cite{amiot:90}, whereas the hyperfine splitting between the different thresholds is $k_B \times 0.33$~K. Each potential has many bound states. For clarity, only one of them is shown in Fig.~\ref{fig-Feshbach}.

The general treatment of scattering resonances given in Sec.~\ref{sec-resonance} is also applicable to Feshbach resonances. In particular, one can show that the $S$-matrix is again given by Eq.~(\ref{eq-S-matrix-single-channel}). To obtain this result, one often uses a formalism based on Green's functions and projection operators introduced by Feshbach \cite{feshbach:58,feshbach:62}, in order to clearly separate the open-channel and closed-channel subspaces. For a discussion of this formalism in the context of ultracold gases see e.g.\ Ref.~\cite{timmermans:99}.

In the collisions of cold alkali atoms, there are typically many Feshbach resonances. The molecular state is a vibrationally highly excited state. For homonuclear molecules, the radiative decay rates into lower vibrational states are negligible due to lack of an electric dipole moment. But in a dense sample, inelastic collisions of the molecule with another atom or molecule can lead to significant rates for vibrational de-excitation. These processes are neglected in this paper.

Feshbach resonances in the collisions of cold atoms differ substantially from most scattering resonances in other fields of physics, insofar as the energy of the molecular state can be tuned noticeably by applying an external magnetic field $B$. The tunability arises from the different magnetic moments of the different channels. If this difference is, say, one Bohr magneton, a magnetic field of 1000~G creates a shift of $k_B \times 0.07$~K. Therefore low temperatures of the incoming flux are needed, otherwise thermal broadening would render this shift insignificant.

Let $B_{res}$ denote the magnetic field, at which the energy of the molecular state matches the open-channel threshold. Then the magnetic-field dependence of the energy $E_{res}$, at which the Feshbach resonance occurs, can be approximated linearly by
\begin{eqnarray}
 \label{eq-E-res}
E_{res}(B) = (B-B_{res}) \Delta \mu
\end{eqnarray}
as long as $|B-B_{res}|$ is not too large. $\Delta \mu$ denotes the difference in the magnetic moments of the two channels. $\Delta \mu$ can be positive or negative. If $B$ approaches $B_{res}$, then $E_{res} \rightarrow 0$ i.e.\ the Feshbach resonance occurs at energy zero. A small correction to Eq.~(\ref{eq-E-res}) is necessary if $B$ is very close to $B_{res}$. This correction is typically of the order of $\Delta B$ (see below). This so-called resonance shift is neglected in the discussion in the present paper. But it is included in our coupled-channels calculations.

For $E_{res}(B)<0$ the molecular state is truly bound, which means that the state cannot decay. When varying $B$ so that $E_{res}$ increases, dissociation abruptly sets in at $E_{res}(B)=0$. Therefore the low-energy edge of the continuum is called dissociation threshold.

An interesting situation arises, if one ramps $B$ through $B_{res}$ in the direction such that $E_{res}$ moves from above to below threshold. This converts the molecular state from quasi-bound to truly bound. If incoming atomic flux is present during the ramp, then population that was transiently in the molecular state during scattering while above threshold, will remain in the state after the threshold is crossed. Thus long-lived molecules can be produced. These can later be dissociated at will by ramping $B$ back through $B_{res}$. This technique was used in several recent experiments to produce ultracold molecules from ultracold atomic gases \cite{regal:03,herbig:03,duerr:04,strecker:03,cubizolles:03,jochim:03a,xu:03,zwierlein:04,thompson:05}. Between the association and the dissociation, the atomic and the molecular cloud can be spatially separated by applying a Stern-Gerlach field.

We emphasize the difference between a Feshbach resonance and a shape resonance. A shape resonance occurs in single-channel scattering. The quasi-bound state is typically localized behind the centrifugal barrier. It could theoretically become a truly bound state by increasing the potential depth, but experimentally, one can usually not tune its energy. A Feshbach resonance is a multi-channel resonance. The quasi-bound state has a spin configuration other than the incoming flux. The energy of the quasi-bound state can be tuned with a magnetic field and this state can become a truly bound state by tuning its energy below the open-channel threshold. Despite their differences, both types of resonances have in common that there is a quasi-bound state above threshold. Moreover, both types of resonances are well described by the same Breit-Wigner expression for the $S$-matrix Eq.~(\ref{eq-S-matrix-single-channel}).

\subsection{Low-Energy Feshbach Resonances}
As discussed in Sec.~\ref{sec-threshold}, $s$-wave scattering usually dominates at low energies. If the magnetic field is held near a Feshbach resonance at $B_{res}$, then the low-energy scattering is affected, of course. According to Eqs.~(\ref{eq-threshold-law}) and (\ref{eq-threshold-Gamma}), $\delta_0$ and $\Gamma$ both vanish like $O(k)$ for $k \rightarrow 0$. One can thus linearize the tangent in Eq.~(\ref{eq-delta-res}), yielding $\delta_0^{res} \sim \tan \delta_0^{res} = \hbar\Gamma/[2(E_{res}-E)]$. The definition of the scattering length Eq.~(\ref{eq-scattering-length}) then yields
\begin{eqnarray}
a = a_{bg} - \frac{\hbar}{2E_{res}} \lim_{k\rightarrow 0}\frac{\Gamma}{k} \; .
\end{eqnarray}
Since $\Gamma \stackrel{k \rightarrow 0}{\sim} O(k)$, this expression is well defined. Inserting Eq.~(\ref{eq-E-res}) for $E_{res}$ one obtains
\begin{eqnarray}
 \label{eq-a-vs-B}
a = a_{bg} \left( 1-\frac{\Delta B}{B-B_{res}} \right) \; ,
\end{eqnarray}
where the (magnetic-field) width of the Feshbach resonance is defined as
\begin{eqnarray}
 \label{eq-Delta-B}
\Delta B = \frac{\hbar}{2 a_{bg}\Delta\mu} \lim_{k\rightarrow 0}\frac{\Gamma}{k} \; .
\end{eqnarray}
$\Gamma$ is always positive, while $\Delta \mu$ and $a_{bg}$ can be positive or negative, independent of each other. Therefore $\Delta B$ can be positive or negative.

Low-energy scattering in the vicinity of a Feshbach resonance can thus be described analytically with only three parameters $a_{bg},B_{res},\Delta B$. The resulting pole and zero in the total cross section $\sigma = 8\pi a^2$ result from interference between background scattered and resonantly scattered wave, just like in the Beutler-Fano profile. Various experiments \cite{inouye:98,roberts:98,stenger:99,cornish:00,loftus:02,volz:03,widera:04} with ultracold atoms observed the behavior predicted by Eq.~(\ref{eq-a-vs-B}).

For sufficiently small $k$, Eq.~(\ref{eq-Delta-B}) can be solved for $\Gamma$. Inserting $E=\hbar^2 k^2 /(2m_{red})$, one obtains the threshold law for the decay rate
\begin{eqnarray}
\Gamma = \frac{2 \Delta B \Delta\mu}{\hbar^2} \; a_{bg} \sqrt{2m_{red}E} \; .
\end{eqnarray}
Recently, this behavior was also experimentally observed with ultracold atoms \cite{mukaiyama:04,duerr:04a}.

\subsection{Selection Rules}
 \label{sec-selection-rules}
As mentioned above, one requirement for a Feshbach resonance is that the interaction Hamiltonian must be able to flip the spins to make transitions between the two relevant channels. This section deals with this issue in more detail.

At large radius, the spins of two colliding ground-state alkali atoms are specified in terms of the hyperfine quantum numbers $|f_1,m_{f1}\rangle$ and $|f_2,m_{f2}\rangle$ of the two atoms. Together with $l,m_l,E$ one obtains a complete set of quantum numbers. The atomic hyperfine spins can be added, yielding the total spin $\vec F = \vec f_1 + \vec f_2$. The corresponding quantum numbers are $F,m_F$.

At shorter radius, the exchange interaction $V_{ex}$ is the dominant term in the interaction Hamiltonian, so that the spins of the valence electrons are coupled to a total electronic spin $\vec S$; and the singlet ($S=0$) and triplet ($S=1$) potentials differ drastically. Hence, the hyperfine quantum numbers $f_1,m_{f1},f_2,m_{f2}$ are not good quantum numbers at short radius. When writing $V_{ex}$ as a matrix in the hyperfine basis, it therefore has large off-diagonal elements, which means that transitions between different hyperfine states are possible. $V_{ex}$ is spherically symmetric and thus conserves $l,m_l$. For incoming $s$ waves, $V_{ex}$ can therefore cause Feshbach resonances only if the molecular state is an $s$-wave state. Since $V_{ex}$ creates only forces internal to the system, the total angular momentum $\vec l +\vec F$ is conserved. Since $m_l$ is conserved, $m_F$ is conserved, too.

In addition, there are much weaker terms in the interaction Hamiltonian. The strongest of these terms is the spin-spin interaction $V_{ss}$, which is the sum of the magnetic dipole-dipole interaction of the valence electrons and the second-order spin-orbit interaction for the valence electrons. $V_{ss}$ can change $l,m_l$ because it is not invariant under spatial rotations. It causes transitions according to the selection rules $\Delta l = 0$ or $\pm2$ and $|\Delta m_l| \leq 2$. For incoming $s$ waves, $V_{ss}$ can therefore cause Feshbach resonances for $d$-wave molecular states. Since $V_{ss}$ is much weaker than $V_{ex}$, the resulting inter-channel coupling is typically also much weaker. Hence, Feshbach resonances caused by $V_{ss}$ are usually much narrower than those caused by $V_{ex}$. Since $V_{ss}$ creates only internal forces, the total angular momentum $\vec l +\vec F$ is again conserved.

Even weaker terms in the Hamiltonian can cause other narrow Feshbach resonances, such as in $^{133}$Cs near 20 Gauss, where an incoming $s$ wave is coupled to a $g$-wave molecular state \cite{mark:05}. $\Delta l$ is always even for the Feshbach resonances in atomic collisions, because the interaction Hamiltonian conserves parity. The only fundamental interaction that does not conserve parity is the weak interaction, but that is negligible here.

When an external magnetic field $\vec B$ is applied, the total angular momentum $\vec l +\vec F$ is no longer conserved, because the external field creates external forces. We consider only the case where $\vec B$ points along the $z$ axis, so that rotational symmetry around the $z$ axis implies the conservation of $m_l+m_F$. Note that if the magnetic field is strong, $f_1,f_2$ are no longer good quantum numbers at large radius.

\section{Resonances with Many Partial Waves}
\label{sec-multichannel}
\subsection{S-Matrix}
In this section, we consider the case where a quasi-bound state couples to unbound states in more than one partial wave. Again, a Breit-Wigner form is obtained for the $S$-matrix (see p.~411 in Ref.~\cite{taylor:72})
\begin{eqnarray}
 \label{eq-S-matrix-multichannel-with-A}
S & = & S^{bg} \left( 1 - \frac{i A}{E-E_{res}+i\hbar\Gamma/2} \right) \; ,
\end{eqnarray}
where $S^{bg}$ and $A$ are matrices and 1 is the identity matrix. While this is often discussed in the context of resonances with couplings between different channels, we here use it for resonances that involve different partial waves. We assume that $S^{bg}$ is diagonal and define $\Gamma_{ll'}=e^{i(\delta_l^{bg} - \delta_{l'}^{bg})} A_{ll'}/\hbar$. We thus obtain
\begin{eqnarray}
 \label{eq-S-matrix-multichannel}
S_{ll'} & = & e^{i(\delta_l^{bg} + \delta_{l'}^{bg})} 
\left( \delta_{ll'} - \frac{i\hbar\Gamma_{ll'}}{E-E_{res}+i\hbar\Gamma/2} \right) .
\end{eqnarray}
Note that this reduces to Eq.~(\ref{eq-S-matrix-single-channel}) in the single partial-wave case, where $\Gamma_{ll'}= \Gamma \delta_{ll'} \delta_{ll_0}$ and $l_0$ denotes the one partial wave that has a resonance.

We assume that there is only one quasi-bound state that causes the resonance. Hence, the matrix $A$ is of rank 1 (see p.~406 in Ref.~\cite{taylor:72}). Combined with unitarity and symmetry of the $S$-matrix, this implies that all $\Gamma_{ll'}$ are real and that they fulfill
\begin{eqnarray}
 \label{eq-product-Gamma}
\Gamma_{ll'}^2 & = & \Gamma_{ll} \, \Gamma_{l'l'} \\
 \label{eq-sum-Gamma}
\Gamma  & = & \sum_l \; \Gamma_{ll} \; .
\end{eqnarray}
We will see in Eq.~(\ref{eq-beta-l}) that the decay rate into the $l$-th partial wave is given by $\Gamma_{ll}$. This quantity cannot be negative. The threshold law Eq.~(\ref{eq-threshold-Gamma}) applies to each diagonal element $\Gamma_{ll}$. The total decay rate $\Gamma$ is the sum of the partial decay rates.

This situation explicitly allows coupling between different partial waves, so that $|S_{ll'}|\neq \delta_{ll'}$. In particular, $|S_{ll}|\neq 1$ and the scattering phases $\delta_l$ can be complex. The $S$-matrix is still unitary and the number of particles is conserved. But the unitarity limit Eq.~(\ref{eq-unitarity-limit}) for $\sigma_l$ known from spherically symmetric potentials can be exceeded here, because flux can be redistributed between partial waves.

When calculating the partial-wave components of the total cross section, one finds an expression that closely resembles a Beutler-Fano profile
\begin{eqnarray}
\label{eq-Fano-multichannel}
\sigma_l & = & \sigma_l^{bg} \; \frac{(\epsilon + {\rm Re} \{ Q\})^2 + (1 + {\rm Im} \{ Q\})^2 }{\epsilon^2 +1}
\end{eqnarray}
with $\epsilon$ and $\sigma_l^{bg}$ from Eqs.~(\ref{eq-def-epsilon}) and (\ref{eq-def-sigma-bg}) and with the complex number 
\begin{eqnarray}
Q = -\; \frac{1}{ \sin \delta_l^{bg} } \sum_{l'} e^{i\delta_{l'}^{bg}} \frac{ \Gamma_{ll'} \sqrt{2l'+1}}{\Gamma \sqrt{2l+1}} \; .
\end{eqnarray}
While $Q$ is a lengthy expression, the key result is that it is independent of $\epsilon$. Hence, $\sigma_l$ is a fairly simple function of energy, namely the product of a parabola and a Lorentzian, just like for the normal Beutler-Fano profile. The full width at half maximum (FWHM) of the Lorentzian as a function of $E$ is $\hbar\Gamma$ just like in the single partial-wave case. But the minimum of $\sigma_l$ can be above zero and the maximum can be above or below the unitarity limit. In the single partial-wave case ($\Gamma_{ll'} = \Gamma \delta_{ll'} \delta_{ll_0}$), one obtains $Q= (-i-\cot \delta_l^{bg}) \delta_{ll_0}$ and Eq.~(\ref{eq-Fano-multichannel}) reduces to the normal Beutler-Fano profile Eq.~(\ref{eq-Fano}).

\subsection{Combination of a Shape Resonance and a Feshbach Resonance in $^{87}$Rb}
\label{sec-two-resonances}
We apply this formalism to $^{87}$Rb with both atoms entering in the hyperfine state $|1,1\rangle$. This is the absolute ground state for atomic $^{87}$Rb so that inelastic two-body collisions cannot occur. This system has a Feshbach resonance at $B_{res}=632.45$~G \cite{marte:02} with $\Delta B= 1.3$~mG \cite{duerr:04a}. The corresponding molecular state is a $d$-wave state, which is coupled to incoming $s$, $d$, and $g$ waves by the spin-spin interaction (see Sec.~\ref{sec-selection-rules}). For the energy range considered here, $g$-wave scattering is negligible. Weaker terms in the Hamiltonian can couple to even higher partial waves, but that is also negligible here.

\begin{figure} [t!]
\includegraphics[width=.4\textwidth]{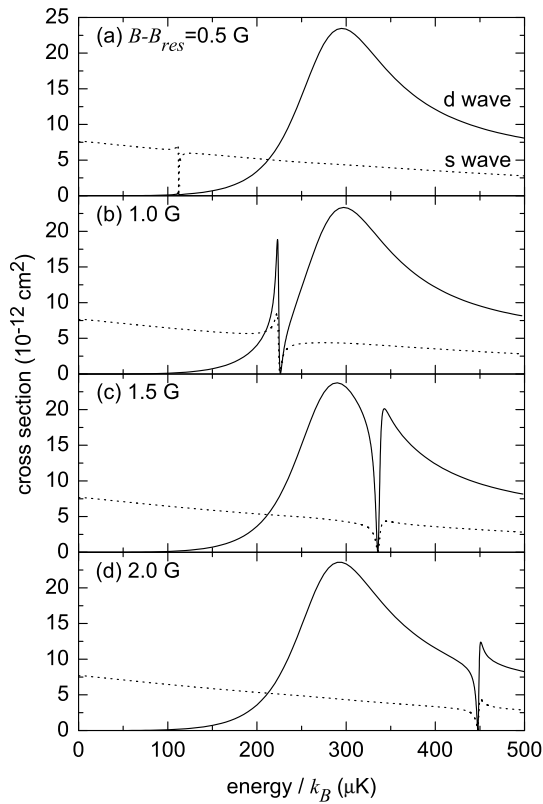}
\caption{\label{fig-shape-and-Feshbach}
Combination of a shape resonance and a Feshbach resonance for scattering of $^{87}$Rb in state $|1,1\rangle$. The partial-wave components $\sigma_l$ of the total cross section are shown for the $s$ wave (dotted line) and the $d$ wave (solid line). The magnetic field $B$ is held at various values above the Feshbach resonance at $B_{res} \sim 632$~G. The Feshbach resonance is much narrower than the shape resonance. By changing $B$, the position of the Feshbach resonance can be tuned through the shape resonance.
 }
\end{figure}

Figure~\ref{fig-shape-and-Feshbach} shows the $s$ and $d$-wave components of the total cross section, $\sigma_0$ and $\sigma_2$ with $m_l=0$ (see below), for different values of the magnetic field $B$. These results were obtained from a coupled-channels calculation. One can clearly see the narrow Feshbach resonance sitting on top of the background, which is modulated due to the broad shape resonance already known from Fig.~\ref{fig-87Rb-shape-resonance}. The energy $E_{res}$, at which the Feshbach resonance occurs, can be tuned with the magnetic field according to Eq.~(\ref{eq-E-res}).

$\Gamma$ in Eq.~(\ref{eq-S-matrix-multichannel}) denotes the total decay rate of the molecular state. The decay rate of the other quasi-bound state that causes the shape resonance does not explicitly occur in Eq.~(\ref{eq-S-matrix-multichannel}), because the open-channel physics including the shape resonance is contained in the energy-dependence of $\delta_l^{bg}$ and $\Gamma_{ll'}$.

When moving $E_{res}(B)$ through the shape resonance, the form of the cross sections $\sigma_l$ near the Feshbach resonance changes, as seen in Fig.~\ref{fig-shape-and-Feshbach}. For $E_{res}(B)<E_{shape}$, the Feshbach resonance increases the cross sections on the low-energy side of the Feshbach resonance and reduces the cross sections on the high-energy side. For $E_{res}(B)>E_{shape}$ this is reversed. This is because when moving $E$ through the shape resonance, $\delta_2^{bg}$ changes by almost $\pi$, as seen in Fig.~\ref{fig-87Rb-shape-resonance}a. This phenomenon is called $q$-reversal, see e.g.\ Ref.~\cite{cornett:99}.

We now discuss, why $m_l=0$ for all outgoing partial waves. As mentioned in Sec.~\ref{sec-selection-rules}, rotational symmetry around $B$, which points along the $z$ axis, implies that $m_l+m_F$ is conserved. Since the incoming channel has $m_F=2, m_l=0$, all outgoing channels must have $m_F + m_l=2$. For energetic reasons, $m_F=2$ is the only possible spin state for the outgoing flux, so that all outgoing waves must have $m_F=2, m_l=0$. Note that {\em during} the collision, couplings to all states with $m_F + m_l=2$ are possible. This is a necessary ingredient, since the molecular state is an almost pure $m_F=4, m_l=-2$ state.

Note that in the case of the above-mentioned resonance near 632~G, $V_{ss}$ is required for the molecules to decay, since $m_l$ has to change by $+2$ for both the outgoing $s$ and $d$ wave, leading to comparable amplitudes for both partial waves. If the molecular state were an $s$-wave state instead, $V_{ss}$ would still couple it to the outgoing $d$ wave, but the much stronger $V_{ex}$ would create a strong coupling to the outgoing $s$ wave, resulting in a very small outgoing $d$-wave fraction. Conversely, if the molecules had $l=2,m_l=0$ they could decay into the outgoing $d$ wave by the strong $V_{ex}$ since $l$ and $m_l$ would not need to change. For decay into the outgoing $s$ wave, the much weaker $V_{ss}$ would be needed as $l$ would need to change by $-2$ and the population of this partial wave would therefore be strongly suppressed as compared to population in the $d$-wave. Such resonances do exist, e.g. in $^{87}$Rb at 551.47~G and 831.29~G \cite{marte:02}. Unfortunately, these two resonances are so narrow ($\Delta B \sim 0.2$~mG each), that creating molecules at these resonances is difficult.

\subsection{Extracting the Partial Decay Rates}
\label{sec-Gamma-matrix}
The $S$-matrix for the above situation was numerically calculated on a fine grid in the $E$-$B$-plane, in order to extract the decay-rate matrix $\Gamma_{ll'}$ and the background scattering-phases $\delta_l^{bg}$ as defined in Eq.~(\ref{eq-S-matrix-multichannel}). The extraction of the $\delta_l^{bg}$ is easy, because they contain only the open-channel physics. This does include the shape resonance, so that the $\delta_l^{bg}$ depend on energy, but it does not include the Feshbach resonance, so that the $\delta_l^{bg}$ depend hardly on the magnetic field. The values of $\delta_l^{bg}(E)$ can therefore simply be read off from the $S$-matrix for pretty much any $B$ sufficiently far away from the Feshbach resonance. Since the open-channel physics is almost independent of $B$, the result for $\delta_l^{bg}$ is essentially the same as shown for $B=0$ in Fig.~\ref{fig-87Rb-shape-resonance}a.

As discussed in the context of Eq.~(\ref{eq-threshold-Gamma}), the decay-rate matrix of the molecular state $\Gamma_{ll'}$ depends on energy. But it depends hardly on the magnetic field. Hence, the extraction of $\Gamma_{ll'}(E)$ from the numerical results for $S(E,B)$ is also fairly easy. To this end, we take the modulus squared of Eq.~(\ref{eq-S-matrix-multichannel}) for $l\neq l'$ and insert Eq.(\ref{eq-E-res}), yielding
\begin{eqnarray}
\lefteqn{ |S_{ll'}|^2 \stackrel{l\neq l'}{=} } && \nonumber \\
&& \frac{\hbar^2 \Gamma_{ll'}^2(E)/\Delta\mu^2 }{(-B+B_{res}+E/\Delta\mu)^2+\hbar^2\Gamma^2(E)/(2 \Delta\mu)^2} \; .
\end{eqnarray}
Considered as a function of $B$ at constant $E$, this is simply a Lorentzian. We fit this to the numerical results for $|S_{02}(E,B)|^2$ at constant $E$ and obtain three fit parameters $|\Gamma_{02}(E)/\Delta\mu|$, $|\Gamma(E)/\Delta\mu|$, and $(B_{res}+E/\Delta\mu)$. A combination of the results of the last fit parameter for various values of $E$ yields $B_{res}=632.3$~G and $\Delta \mu = k_B \times 224~\mu$K/G $=3.33~\mu_B$ with the Bohr magneton $\mu_B$. The deviation between the theoretical and experimental value for $B_{res}$ is no problem, as long as the comparison between experiment and theory is performed in terms of $B-B_{res}$.

\begin{figure} [t!]
\includegraphics[width=.4\textwidth]{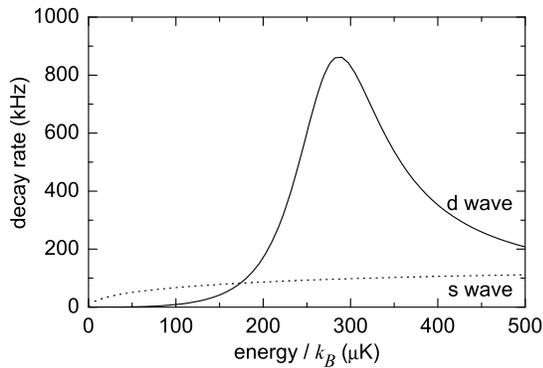}
\caption{\label{fig-decay-rates}
Decay rates of the molecular state for the 632-G Feshbach resonance in $^{87}$Rb. The partial decay rates into the $s$ wave (dotted line) and $d$ wave (solid line) are shown. The $d$-wave shape resonance obviously has a drastic effect on the $d$-wave decay rate.
 }
\end{figure}

Knowing $\Delta\mu$, one obtains $|\Gamma_{02}(E)|$ and $\Gamma(E)$. Using Eqs.~(\ref{eq-product-Gamma}) and (\ref{eq-sum-Gamma}), one easily obtains $\Gamma_{00}(E)$ and $\Gamma_{22}(E)$, except for the ambiguity of which is which. This can easily be resolved by inspection of the diagonal elements of $S(E,B)$. In addition, inspection of the off-diagonal elements of $S(E,B)$ yields the sign of $\Gamma_{02}(E)$, which is
\begin{eqnarray}
\label{eq-Gamma-02}
\Gamma_{02}=\Gamma_{20}=-\sqrt{\Gamma_{00}\Gamma_{22}}
\end{eqnarray}
for all $E$ in the present calculation. Thus, the complete decay-rate matrix $\Gamma_{ll'}(E)$ is extracted. The partial decay rates $\Gamma_{00}(E)$ and $\Gamma_{22}(E)$ of the molecular state are shown in Fig.~\ref{fig-decay-rates}. 

\begin{figure} [t!]
\includegraphics[width=.4\textwidth]{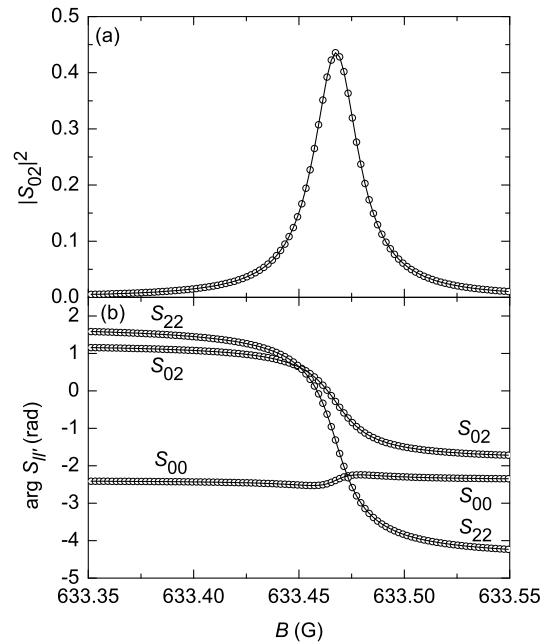}
\caption{\label{fig-fit-S-matrix}
Fitting to the $S$-matrix elements. Results from the coupled-channels calculation (circles) are shown versus magnetic field $B$ for a fixed energy of $k_B \times 255 \ \mu$K. The fit curves (solid lines) are hardly visible, because they agree so well with the coupled-channels results.
 }
\end{figure}

In order to demonstrate the quality of the fit, the results of the coupled-channels calculation are compared to the fit curves in Fig.~\ref{fig-fit-S-matrix}. Parts (a) and (b) show modulus squared and phase of the $S$-matrix elements, respectively. For clarity, some quantities that are trivially related to the others are not shown. Symmetry and unitarity of the $S$ matrix imply $S_{20} = S_{02}$, $|S_{00}|^2 = |S_{22}|^2 = 1 - |S_{02}|^2$, and $\arg S_{02} = (\pi +\arg S_{00} + \arg S_{22})/2$. All curves shown in Fig.~\ref{fig-fit-S-matrix} are well described by only four parameters $\delta_0^{bg}$, $\delta_2^{bg}$, $\Gamma_{02}$, and $\Gamma$. The excellent agreement between the fit and the coupled-channels results demonstrates that the Breit-Wigner form Eq.~(\ref{eq-S-matrix-multichannel}) is a very good approximation.

\section{Molecule Dissociation}
\label{sec-dissociation} 
\subsection{General}
In this section, we discuss the decay of the molecular state as observed in Ref.~\cite{volz:cond-mat/0410083}. As already described at the end of Sec.~\ref{sec-Feshbach-general}, molecules are formed by ramping $B$ through $B_{res}$. A Stern-Gerlach field then removes the incoming flux from the spatial region of interest and finally the molecules are dissociated by jumping $B$ above $B_{res}$ and holding it there at a fixed value. For further experimental details, see Refs.~\cite{duerr:04,volz:cond-mat/0410083}.

Due to conservation of energy during the decay, the {\em mean} energy of the outgoing wave is given by 
\begin{eqnarray}
E = E_{res}(B)
\end{eqnarray}
with $E_{res}$ from Eq.~(\ref{eq-E-res}). Due to the finite lifetime of the molecular state, the energy of the outgoing wave has a width of $\hbar \Gamma$.

Note the difference between this dissociation experiment and time-independent elastic scattering: for time-independent scattering, the energy of the outgoing wave must be identical to the energy of the incoming wave, whereas in the decay experiment described here the energy of the outgoing wave is adjusted with $B$. In the decay experiment, the energy of the outgoing wave of up to $k_B \times 500~\mu$K is typically much larger than the energy of the incoming wave with $E < k_B \times 1~\mu$K, from which the molecules were originally formed. This is possible because the magnetic-field ramp makes the Hamiltonian explicitly time dependent, so that energy is not conserved in a simple-minded fashion. Ultimately, the additional energy in the outgoing wave comes from the power supply that creates the magnetic-field ramp.

In the decay experiment, the original incoming wave has such low energy, that only incoming $s$-waves are relevant. This implies that $m_l=0$ for the incoming wave. As discussed in Sec.~\ref{sec-two-resonances}, this implies that all outgoing partial waves must also have $m_l=0$. And again, only $s$ and $d$ waves are important for the dissociation in the energy range considered here.

\subsection{Connection between Scattering and Decay}
The decay experiment is closely related to the scattering experiment discussed in Secs.~\ref{sec-two-resonances} and \ref{sec-Gamma-matrix}. The link can be established by inspection of the scattering wave function, which is obtained from the combination of Eqs.~(\ref{eq-define-S-matrix}) and (\ref{eq-sum-S-matrix})
\begin{eqnarray}
\lefteqn{ \psi^{(+)}_{l'}(\vec r) \stackrel{r \rightarrow \infty}{\sim} 
(-1)^{l'} \frac{e^{-ikr}}{r} \; Y_{l'0} (\vartheta)   }&& \nonumber \\
&& - \frac{e^{ikr}}{r} \; \sum_{l=0}^\infty S_{ll'}^{bg} Y_{l0} (\vartheta) 
- \frac{e^{ikr}}{r} \; \sum_{l=0}^\infty S_{ll'}^{res} Y_{l0} (\vartheta) \; .
\end{eqnarray}
The first term is the incoming wave, the second term is the background scattered wave and the third term is the resonantly scattered wave. The resonantly scattered part consists of population that made the transition to the molecular state and then decayed back to the open channel. In the decay experiment in Ref.~\cite{volz:cond-mat/0410083}, the Stern-Gerlach field removed the incoming wave and along with it the background scattered wave. Hence, these two terms must be removed to describe the decay experiment
\begin{eqnarray}
\label{eq-psi-decay}
\psi_{decay}(\vec r) \stackrel{r \rightarrow \infty}{\sim} 
- \frac{e^{ikr}}{r} \; \sum_{l=0}^\infty S_{ll'}^{res} Y_{l0} (\vartheta) \; .
\end{eqnarray}
This crucial step makes the connection between scattering and dissociation.

From Eq.~(\ref{eq-S-matrix-multichannel}) one obtains
\begin{eqnarray}
S_{ll'}^{res} = - e^{i(\delta_l^{bg}+\delta_{l'}^{bg})} 
\frac{i\hbar\Gamma_{ll'}}{E-E_{res} +i\hbar\Gamma/2}
\end{eqnarray}
Inserting this and Eq.~(\ref{eq-Gamma-02}) into Eq.~(\ref{eq-psi-decay}) and choosing $l'=0$ yields
\begin{eqnarray}
\psi_{decay}(\vec r) && \stackrel{r \rightarrow \infty}{\sim} 
\frac{e^{ikr}}{r} \; \frac{i\hbar e^{i\delta_0^{bg}} \sqrt{\Gamma_{00}}}{E-E_{res} +i\hbar\Gamma/2} \nonumber \\
&& \times \left(e^{i\delta_0^{bg}} \sqrt{\Gamma_{00}} Y_{00} - e^{i\delta_2^{bg}} \sqrt{\Gamma_{22}} Y_{20}(\vartheta) \right) \; . \quad \quad 
\end{eqnarray}
We abbreviate
\begin{eqnarray}
 \label{eq-beta-l}
\beta_l & = & \frac{\Gamma_{ll}}{\Gamma} \\
\delta_{rel} & = & \delta_2^{bg} - \delta_0^{bg}
\end{eqnarray}
and obtain
\begin{eqnarray}
\psi_{decay}(\vec r) \stackrel{r \rightarrow \infty}{\sim} \! 
\tilde g(r,E) \! \left(\sqrt{\beta_0} Y_{00} - e^{i\delta_{rel}} \sqrt{\beta_2} Y_{20}(\vartheta) \right)
\end{eqnarray}
where $\tilde g(r,E)$ is the radial wave function. The time-domain version thereof $g(r,t)$ is related to $\tilde g(r,E)$ by a Fourier transform.

Equation (\ref{eq-sum-Gamma}) implies that $\beta_0+\beta_2=1$ so that $\beta_l$ is the branching ratio for decay into the $l$-th partial wave. $\delta_{rel}$ is the relative phase between the two partial waves. The above definitions of $\beta_l$ and $\delta_{rel}$ do not explicitly depend on $E$. But they do involve $\Gamma_{ll'}(E)$ and $\delta_l^{bg}(E)$. The energy-dependence of these quantities within the width of the resonance $\hbar \Gamma$ is negligible, so that one can simply evaluate these quantities at $E=E_{res}(B)$.

In conclusion, we showed how the dissociation of ultracold molecules into more than one partial wave is related to a scattering experiment. This makes it possible to use coupled-channels calculations for scattering experiments to analyze dissociation experiments, such as the one in Ref.~\cite{volz:cond-mat/0410083}. The analysis in Ref.~\cite{volz:cond-mat/0410083} shows that the experiment agrees well with the theory described here.

\acknowledgments
S.K.\ acknowledges support from the Netherlands Organization for Scientific Research (NWO). E.K.\ and B.V.\ acknowledge support from the Stichting FOM, financially supported by NWO.

\bigskip

\appendix

\section{Poles of the S-Matrix}
\label{sec-app-poles-of-S}
\subsection{Analytic Continuation}
Insight into scattering resonances can be gained in a very general formalism that makes no use of the specific form of the potential. The starting point is the observation that the $S$-matrix is typically an analytic function of $k$. The key idea is then to consider the analytic continuation of $S(k)$ into the complex $k$-plane. One can show that this continuation is unique, but cannot always cover the whole complex plane. The physical meaning of $k$ as the magnitude of the wave vector requires it to be real and non-negative. But the continuation into the complex plane will offer additional physical insight, as we will see in the following. For simplicity, we consider only a spherically symmetric single-channel potential in this appendix. More details about the topics discussed in this appendix can be found in Ref.~\cite{taylor:72}.

In this discussion, one usually considers the Jost function $\tilde f_l(k)$ instead of the $S$-matrix. Like the $S$-matrix, the Jost function is also defined by the coefficients in the scattering state Eq.~(\ref{eq-define-S-matrix}), but with a different normalization
\begin{eqnarray}
\label{eq-def-Jost}
\phi_l(r) \stackrel{r \rightarrow \infty}{\sim} 
\left( (-1)^l \tilde f_l(k) \frac{e^{-ikr}}{r} - \tilde f_l^*(k^*) \frac{e^{ikr}}{r} \right) Y_{l0}(\vartheta) \quad
\end{eqnarray}
The Jost function also has a unique analytic continuation into the complex $k$ plane. For most potentials, $\tilde f_l(k)$ is analytic everywhere in this plane, except for the negative imaginary axis. Note that $\tilde f_l(k)$ and $\tilde f_l^*(k^*)$ cannot both vanish for the same value of $k$. As a consequence, $\tilde f_l(k)$ cannot vanish if $k$ is real.

The $S$-matrix is easily obtained from the Jost function:
\begin{eqnarray}
\label{eq-S-Jost}
S_{ll'}(k) = \frac{\tilde f_l^*(k^*)}{\tilde f_l(k)} \; \delta_{ll'} \; .
\end{eqnarray}
The Kronecker symbol comes from our assumption of spherical symmetry. If $k$ is real, this implies $|S_{ll'}|^2=\delta_{ll'}$, i.e.\ unitarity. 

Note that the $S$-matrix has a pole at $k$ if $\tilde f_l(k)=0$. In other words, poles of the $S$-matrix correspond to zeros of the Jost function. If $k$ is real, the Jost function cannot vanish, so that all zeros of the Jost function must lie either in the upper half-plane, i.e.\ ${\rm Im}\{k\}>0$, or in the lower half-plane, i.e.\ ${\rm Im}\{k\}<0$. We will now discuss what the physical meaning of the zeros of the Jost function is. This depends on the half plane, in which the zero is.

\subsection{Bound States}
For complex values of $k$, Eq.~(\ref{eq-def-Jost}) has one term that increases exponentially for large $r$, and one term that decreases exponentially. We assume that there is a point $k_0$ in the {\em upper} half-plane with $\tilde f_l(k_0)=0$. Here, Eq.~(\ref{eq-def-Jost}) has only one term. This term falls off exponentially, because we assumed that $k_0$ is in the upper half-plane. Hence, Eq.~(\ref{eq-def-Jost}) is a normalizable eigenstate of the Hamiltonian. In other words, this is a bound state.

The bound-state energy is $\hbar^2 k_0^2 /(2m_{red})$. Since the Hamiltonian is hermitian, this energy must be real. Since $k_0$ is in the upper half-plane, it follows that $k_0=i\alpha$ with $\alpha$ real and positive. The bound-state energy is then $- \hbar^2 \alpha^2 /(2m_{red})$. Conversely, one can show that if the Hamiltonian has a bound state with angular momentum $l$ and energy $-\hbar^2 \alpha^2 /(2m_{red})$, then $\tilde f_l(i \alpha)=0$.

To summarize, bound states have a one-to-one correspondence to poles of the $S$-matrix in the upper half of the complex $k$ plane. All these poles must lie on the imaginary axis.

\subsection{Resonances}
We now investigate what happens if the Jost function vanishes at a point $k_0$ in the {\em lower} half-plane. Again, the first term in Eq.~(\ref{eq-def-Jost}) vanishes, but this does not result in a normalizable eigenstate, because the remaining term {\em increases} exponentially for large $r$. Unlike before, $k_0$ does not have to lie on the imaginary axis.

The Jost function can thus have a large number of zeros everywhere in the lower half-plane. Most of them are usually uninteresting. The only interesting ones are those, which lie close to the positive real axis because they can create resonances, as we will show now. The correspondence between resonances and zeros of the Jost function in the lower half-plane is not so clear-cut one-to-one as in the case of bound states. Still, when considering only those poles that lie close to the positive real axis and when ignoring some rather special cases (see p.~241 in Ref.~\cite{taylor:72}), one can think of this correspondence as being one-to-one.

The linear approximation further below is usually made in terms of energy, rather than $k$. Obviously, one can substitute $E = \hbar^2 k^2/(2 m_{red})$ and obtain $\tilde f_l(E)$. Care must be taken, because the transition from $k$ to $E$ is a two-to-one mapping, so that $\tilde f_l(E)$ is a function on a two-sheeted Riemann surface. ${\rm Im}\{k\}>0$ corresponds to the first sheet of $E$ (also called physical sheet) and ${\rm Im}\{k\}<0$ corresponds to the second (or unphysical) sheet of $E$.

If $\tilde f_l(k_0)=0$ with $k_0$ in the lower half-plane, then $\tilde f_l(E)$ has a zero on the unphysical sheet at $E_0 =\hbar^2 k_0^2 / (2m_{red})$. We assume that $\tilde f_l(E)$ has a {\em simple} zero at $E_0$, so that near $E_0$ we can use a linear approximation
\begin{eqnarray}
\label{eq-Jost-approx}
\tilde f_l(E) \approx c (E-E_0)
\end{eqnarray}
with a nonzero value of $c=\frac{d\tilde f_l}{dE} \big|_{E_0}$. We already mentioned that $k_0$ must lie just slightly below the positive real $k$ axis to create a noticeable resonance. This implies that $E_0$ also lies just slightly below the positive real $E$ axis. Hence, there is some interval on the positive real $E$ axis, where Eq.~(\ref{eq-Jost-approx}) is a good approximation. From Eq.~(\ref{eq-S-Jost}), one obtains
\begin{eqnarray}
S_{ll} \approx \frac{c^*(E-E_0)^*}{c\, (E-E_0)}
\end{eqnarray}
for $E$ in this interval on the real axis. We abbreviate the prefactor that is independent of $E$ as
\begin{eqnarray}
S_{ll}^{bg} = \frac{c^*}{c}
\end{eqnarray}
and we split $E_0$ into its real and imaginary part
\begin{eqnarray}
E_0 = E_{res} - i \frac{\hbar\Gamma}{2} \; .
\end{eqnarray}
Here, $E_{res}$ and $\Gamma$ must be positive, because we assumed that the zero of the Jost function lies slightly below the positive real $k$ axis. We thus obtain the Breit-Wigner expression
\begin{eqnarray}
S_{ll} \approx S_{ll}^{bg} \left( 1 - \; \frac{i\hbar \Gamma}{E-E_{res} + i\hbar \Gamma/2} \right) \; .
\end{eqnarray}
The real and imaginary parts of $E_0$ are thus identified as the position and the width of the resonance.

With this approach to the Breit-Wigner form, $S_{ll}^{bg}$ and $\Gamma$ are independent of $E$. If the resonance is narrow, i.e.\ $\Gamma$ is small, then this is a good approximation. Broader resonances can be included in the formalism by allowing $S_{ll}^{bg}$ and $\Gamma$ to depend on $E$. For a very broad resonance, the background scattering phase can change considerably over the width of the resonance, and it becomes questionable whether one really should regard this as a resonance.


\end{document}